\documentstyle[12pt]{article}
\def\cn{\hbox{cn}}
\def\sn{\hbox{sn}}

\def\emline#1#2#3#4#5#6{%
       \put(#1,#2){\special{em:moveto}}%
       \put(#4,#5){\special{em:lineto}}}
\def\newpic#1{}

\def\bea{\begin{eqnarray}}
\def\eea{\end{eqnarray}}

\def\beq{\begin{equation}}
\def\eeq{\end{equation}}
\def\ba{\beq\new\begin{array}{c}}
\def\ea{\end{array}\eeq}
\def\be{\ba}
\def\ee{\ea}

\makeatletter
\newdimen\normalarrayskip 
\newdimen\minarrayskip 
\normalarrayskip\baselineskip
\minarrayskip\jot
\newif\ifold \oldtrue \def\new{\oldfalse}
\def\arraymode{\ifold\relax\else\displaystyle\fi} 
\def\eqnumphantom{\phantom{(\theequation)}} 
\def\@arrayskip{\ifold\baselineskip\z@\lineskip\z@
\else
\baselineskip\minarrayskip\lineskip2\minarrayskip\fi}
\def\@arrayclassz{\ifcase \@lastchclass \@acolampacol \or
\@ampacol \or \or \or \@addamp \or
\@acolampacol \or \@firstampfalse \@acol \fi
\edef\@preamble{\@preamble
\ifcase \@chnum
\hfil$\relax\arraymode\@sharp$\hfil
\or $\relax\arraymode\@sharp$\hfil
\or \hfil$\relax\arraymode\@sharp$\fi}}
\def\@array[#1]#2{\setbox\@arstrutbox=\hbox{\vrule
height\arraystretch \ht\strutbox
depth\arraystretch \dp\strutbox
width\z@}\@mkpream{#2}\edef\@preamble{\halign
\noexpand\@halignto
\bgroup \tabskip\z@ \@arstrut \@preamble \tabskip\z@ \cr}%
\let\@startpbox\@@startpbox \let\@endpbox\@@endpbox
\if #1t\vtop \else \if#1b\vbox \else \vcenter \fi\fi
\bgroup \let\par\relax
\let\@sharp##\let\protect\relax
\@arrayskip\@preamble}
\def\eqnarray{\stepcounter{equation}%
\let\@currentlabel=\theequation
\global\@eqnswtrue
\global\@eqcnt\z@
\tabskip\@centering
\let\\=\@eqncr
$$%
\halign to \displaywidth\bgroup
\eqnumphantom\@eqnsel\hskip\@centering
$\displaystyle \tabskip\z@ {##}$%
\global\@eqcnt\@ne \hskip 2\arraycolsep
$\displaystyle\arraymode{##}$\hfil
\global\@eqcnt\tw@ \hskip 2\arraycolsep
$\displaystyle\tabskip\z@{##}$\hfil
\tabskip\@centering
&{##}\tabskip\z@\cr}
\begingroup\ifx\undefined\newsymbol \else\def\input#1 {\endgroup}\fi
\newfont{\hr}{msbm10}
\newfont{\ams}{msam10}

\font\numbers=cmss12
\font\upright=cmu10 scaled\magstep1
\def\stroke{\vrule height8pt width0.4pt depth-0.1pt}
\def\topfleck{\vrule height8pt width0.5pt depth-5.9pt}
\def\botfleck{\vrule height2pt width0.5pt depth0.1pt}
\def\Zmath{\vcenter{\hbox{\numbers\rlap{\rlap{Z}\kern 0.8pt\topfleck}\kern
2.2pt \rlap Z\kern 6pt\botfleck\kern 1pt}}}
\def\Qmath{\vcenter{\hbox{\upright\rlap{\rlap{Q}\kern
3.8pt\stroke}\phantom{Q}}}}
\def\Nmath{\vcenter{\hbox{\upright\rlap{I}\kern 1.7pt N}}}
\def\Cmath{\vcenter{\hbox{\upright\rlap{\rlap{C}\kern
3.8pt\stroke}\phantom{C}}}}
\def\Rmath{\vcenter{\hbox{\upright\rlap{I}\kern 1.7pt R}}}
\def\Z{\ifmmode\Zmath\else$\Zmath$\fi}
\def\Q{\ifmmode\Qmath\else$\Qmath$\fi}
\def\N{\ifmmode\Nmath\else$\Nmath$\fi}
\def\C{\ifmmode\Cmath\else$\Cmath$\fi}
\def\R{\ifmmode\Rmath\else$\Rmath$\fi}

\newcounter{app}

\def\app{\setcounter{equation}{0}
\def\theequation{\Alph{app}.\arabic{equation}}\par
\addvspace{4ex}
\@afterindentfalse
\secdef\@app\@dapp}

\newcommand\@app{\@startsection {app}{1}{0ex}%
{-3.5ex \@plus -1ex \@minus -.2ex}%
{2.3ex \@plus.2ex}%
{\normalfont\Large\bf}}
\def\@dapp#1{%
{\parindent \z@ \raggedright \bf #1}\par\nobreak}
\def\l@app#1#2{\ifnum \c@tocdepth >\z@
\addpenalty\@secpenalty
\addvspace{1.0em \@plus\p@}%
\setlength\@tempdima{8em}%
\begingroup
\parindent \z@ \rightskip \@pnumwidth
\parfillskip -\@pnumwidth
\leavevmode \bfseries
\advance\leftskip\@tempdima
\hskip -\leftskip
#1\nobreak\hfil \nobreak\hb@xt@\@pnumwidth{\hss #2}\par
\endgroup\fi}
\newcounter{sapp}[app]

\def\sapp{\def\theequation{\Alph{app}.\arabic{equation}}
\par
\@afterindentfalse
\secdef\@sapp\@dsapp}
\newcommand{\@sapp}{\@startsection{sapp}{2}{\z@}%
{-3.25ex\@plus -1ex \@minus -.2ex}%
{1.5ex \@plus .2ex}%
{\normalfont\large\bfseries}}

\def\@dsapp#1{%
{\parindent \z@ \raggedright \bf #1}\par\nobreak}
\newcommand{\l@sapp}{\@dottedtocline{2}{1.5em}{2.3em}}

\def\2{{1\over 2}}
\def\N2{${\cal N}=2$}

\def\be{ \begin{eqnarray} }
\def\ee{ \end{eqnarray} }

\def\bea{\begin{eqnarray}}
\def\eea{\end{eqnarray}}

\def\beq{\begin{equation}}
\def\eeq{\end{equation}}
\def\ba{\beq\new\begin{array}{c}}
\def\ea{\end{array}\eeq}
\def\be{\ba}
\def\ee{\ea}

\def\R{Ruijsenaars\ }

\begin{document}

\setcounter{footnote}{1}
\def\thefootnote{\fnsymbol{footnote}}
\begin{center}
\hfill FIAN/TD-02/00,\ ITEP/TH-05/00\\
\hfill hep-th/0001168\\
\vspace{0.3in}
{\large\bf Double Elliptic Systems: Problems and
Perspectives
\footnote{Talk presented at the International Workshop
"Supersymmetries and quantum symmetries".}
}
\end{center}
\centerline{{\large A.Mironov}\footnote{Theory Dept.,
Lebedev Physical Inst. and ITEP, Moscow,
Russia}, {\large A.Morozov}\footnote{ITEP, Moscow,
Russia}}

\bigskip

\abstract{\footnotesize
This talk presents a list of problems related to the
double-elliptic (Dell) integrable systems with
elliptic dependence on both momenta and coordinates.
As expected, in the framework of Seiberg-Witten (SW)
theory the recently discovered explicit self-dual family
of 2-particle Dell Hamiltonians
provides a perturbative period matrix which is
a logarithm of the ratio of the (momentum-space)
theta-functions.
}

\begin{center}
\rule{5cm}{1pt}
\end{center}

\bigskip
\setcounter{footnote}{0}

Double-elliptic integrable systems for a long time remain among
the most mysterious objects of string and mathematical physics:
their existence is implied by many branches
of science, while they have been never explicitly constructed.

Recently we have finally obtained \cite{BMMM3,MM} an explicit expression
for the Dell Hamiltonian in the
simplest case of $SU(2)$ (two-particle model) and suggested a direct way to
find -- at least in a somewhat transcendental form -- the Hamiltonians for
all the $SU(N)$.
It makes sense now to recollect all the subjects where the
Dell systems were expected to manifest themselves and begin the program of
putting things together. This note presents just a list of topics and open
problems with minimal comments.

\subsection*{1\hspace{2mm} Dell system}

According to \cite{BMMM3,MM} the double elliptic system of
2 particles in the center-of-mass frame is described by the Hamiltonian:

\be\label{dellH}
H(p,q|k,\tilde k) =
\alpha(q|\tilde k)
\cn\left(p\beta(q|k,\tilde k)) \bigg|\
\frac{k\alpha(q|\tilde k)} {\beta(q|k,\tilde k)}
\right).
\ee
Here

\be
\alpha^2(q|\tilde k) \equiv 1 -
\frac{2g^2}{\sn^2(q|\tilde k)}, \ \ \
\beta^2(q|k,\tilde k) \equiv k'^2 + k^2\alpha^2(q|\tilde k) = 1 -
\frac{2g^2 k^2}{\sn^2(q|\tilde k)}
\ee
$\sn$ and $\cn$ are the standard Jacobi functions.
The model is parametrized by two independent (momentum and coordinate) {\it
bare} elliptic curves with elliptic moduli $k$ and $\tilde k$ ($k' =
\sqrt{1-k^2}$ and $\tilde k' = \sqrt{1 - \tilde k^2}$ are the complimentary
elliptic moduli).  The {\it full} spectral curve

\be\label{sc}
H(\eta,\xi|k,\tilde k) = u \ \ \ \ \left(\ = \cn(Q|k)\ \right)
\ee
is characterized by the effective elliptic moduli

\be
k_{eff} = \frac{k \alpha(q|\tilde k)} {\beta(q|k,\tilde k)},\ \ \ \
\tilde k_{eff} = \frac{\tilde k \alpha(q|k)}
{\beta(q|\tilde k,k)}
\ee
Coordinate-momentum duality \cite{duality} interchanges
$k \leftrightarrow \tilde k$, $k_{eff} \leftrightarrow \tilde k_{eff}$
(and $q,p \leftrightarrow Q,P$).

In general $SU(N)$ case, the model describes an interplay between
the four tori: the two bare elliptic curves and two
effective Jacobians of complex dimension $g = N-1$.

The $SU(2)$ Hamiltonian (\ref{dellH}) has a rather nice form in terms
of the $p,q$ variables or their duals $P,Q$.
However, the general construction of
refs.\cite{BMMM3,MM} for $SU(N)$ describes Hamiltonians as ratios
of genus $g$ theta-functions which depend on another kind of
canonical variables -- angle-action variables $p^{Jac}_i, a_i$.
The {\it flat} moduli $a_i$ play the central role in the
SW theory, while $Q_i$ are rather generalizations
of the algebraic moduli; in the most familiar case of the Toda chain
the spectral curve is
$$
w + \frac{\Lambda_{QCD}^{2N}}{w} = \prod_{i=1}^N (\lambda - Q_i)
$$
while
$a_i = \oint_{A_i} \lambda\frac{dw}{w}$. For our Dell system
these variables $a_i$ can be calculated as A-periods

\be\label{a}
a_i=\oint_{A_i} dS
\ee
of the generating differential $dS=\eta d\xi$ on the spectral curve
(\ref{sc}), while its B-periods define the prepotential ${\cal F}(a)$ through

\be\label{F}
{\partial {\cal F}\over\partial a_i}=\oint_{B_i} dS
\ee

It is an interesting open problem to express the Hamiltonians
in terms of $P_i$ and $Q_i$, perhaps, they can acquire a more
transparent form, like it happens for $SU(2)$.
Moreover, in \cite{BMMM3,MM} the Hamiltonians for $N>2$
were found only in the limit $\tilde k=0$ (elliptic-trigonometric
model, the dual of elliptic Ruijsenaars): elliptization of
coordinates (switching on $k \neq 1$) should be straightforward
(the real problem have been to elliptize the momentum dependence),
but still remains to be performed. Another option is to switch to the
"separated" variables ${\cal P}_i$, ${\cal Q}_i$ \cite{sv} such that
$\oint_{A_j}{\cal P}_id{\cal Q}_i=a_i\delta_{ij}$ (while generically
$a_i=\oint_{A_j}\sum_i P_idQ_i$). Generically, they are different from $P_i$,
$Q_i$ but coincide with $P$, $Q$ in the case of $SU(2)$.

To conclude our description of the $SU(2)$ Dell system, we calculate the
perturbative prepotential ({\it i.e.} the leading order of the expansion in
powers of $\tilde k$ when the bare spectral torus degenerates into
sphere) that allows one to establish the identification with physical
theories. In the
forthcoming calculation we closely follow the line of \cite{BMMM2} where the
very similar calculation has been done for the Ruijsenaars model.

When $\tilde k\to 0$, $\sn(q|\tilde k)$ degenerates into the ordinary sine.
For further convenience, we shall
parametrize the coupling constant $2g^2\equiv\sn^2(\epsilon|k)$. Now the
spectral curve (\ref{sc}) acquires the form

\be\label{psc}
\alpha(\xi)\equiv\sqrt{1-{\sn^2(\epsilon|k)\over\sin^2\xi}}=
{u\over \cn
\left(\eta\beta\bigg|k_{eff}\right)}
\ee
Here the variable $\xi$ lives in the cylinder produced after degenerating
the bare coordinate torus. So does the variable $x=1/\sin^2\xi$. Note
that the A-period of the dressed torus shrinks on the sphere to a contour
around $x=0$. Similarly, B-period can be taken as a contour passing from
$x=0$ to $x=1$ and back.

The next step is to calculate variation of the generating differential
$dS=\eta d\xi$ w.r.t. the modulus
$u$ in order to obtain a holomorphic differential:
\be
dv=\left(-i\sn(\epsilon|k)\sqrt{k'^2+k^2u^2}\right)^{-1}
{dx\over x\sqrt{(x-1)(U^2-x)}}
\ee
where $U^2\equiv{1-u^2\over\sn^2(\epsilon|k)}$.
Since

\be
{\partial a\over\partial u}=\oint_{x=0}{\partial dS\over\partial u}
=\oint_{x=0}dv=
-{1\over \sqrt{(1-u^2)(k'^2+k^2u^2)}}
\ee
we deduce that $u=\cn (a|k)$ and $U={\sn(a|k)\over\sn(\epsilon|k)}$. The
ratio of the B- and A-periods of $dv$
gives the period matrix
\be
T={U\over\pi}\int_0^1 {dx\over x\sqrt{(x-1)(U^2-x)}}=
-{1\over i\pi}\lim_{\kappa\to 0}\left(\log{\kappa\over 4}\right)+
{1\over i\pi}\log{U^2\over 1-U^2}
\ee
where $\kappa$ is a small-$x$ cut-off. The $U$ dependent part of this
integral is finite and can be
considered as the \lq\lq true" perturbative correction, while
the divergent part just renormalizes the bare  \lq\lq classical"
coupling constant $\tau$, {\it i.e.} classical part of the prepotential
(see \cite{BMMM2} for further details).
Therefore, the perturbative period matrix is finally
\be
T_{finite}={i\over\pi}\log{\sn^2(\epsilon|k)-
\sn^2 (a|k)\over\sn^2 (a|k)}+\ const\
\longrightarrow
{i\over\pi}\log{\theta_1(a+\epsilon)\theta_1(a-\epsilon)\over\theta_1^2(a)}
\ee
and the perturbative prepotential is the elliptic tri-logarithm (cf.
\cite{6d}). Remarkably, it lives on
the {\it bare} momentum torus, while the modulus of the perturbative curve
(\ref{psc}) is the dressed one.

\subsection*{2\hspace{2mm} Dell systems in the context of modern theory}

\begin{figure}[p]
\special{em:linewidth 0.4pt}
\unitlength 1.00mm
\linethickness{0.4pt}
\begin{picture}(154.67,151.67)
\linethickness{3pt}
\put(64.33,96.67){\framebox(21.00,15.00)[cc]{\parbox{.14\linewidth}{periodic
Toda chain}}}
\put(64.33,78.00){\framebox(21.00,15.00)[cc]{\parbox{.14\linewidth}{elliptic
Calogero model}}}
\put(64.00,58.67){\framebox(25.00,15.00)[cc]
{\parbox{.16\linewidth}{elliptic Ruijsenaars model}}}
\put(64.00,41.67){\framebox(21.33,11.33)[cc]{\parbox{.15\linewidth}{Dell
system}}}
\linethickness{2pt}
\put(64.67,21.00){\framebox(18.67,12.00)[cc]{\parbox{.12\linewidth}{matrix
models}}}
\put(64.67,2.33){\framebox(17.67,10.00)[cc]{\parbox{.12\linewidth}{duality}}}
\put(74.17,76.83){\thicklines\oval(37.00,72.33)[]}
\linethickness{.4pt}
\put(24.33,16.00){\framebox(30.00,16.33)[cc]
{\parbox{.20\linewidth}{$P$, $Q$ vs. $p$, $a$ vs. separated variables}}}
\linethickness{2pt}
\put(113.33,26.00){\framebox(36.33,20.33)[cc]
{\parbox{.25\linewidth}{Hamiltonians as Casimirs: group theory, zonal spherical
functions}}}
\linethickness{.4pt}
\put(113.33,52.67){\framebox(25.33,19.00)[cc]
{\parbox{.17\linewidth}
{zeroes of $\tau$-functions and solitonic solutions
}}}
\put(113.33,76.00){\framebox(35.33,19.33)[cc]
{\parbox{.25\linewidth}
{free-field ({\it e.g.} fermionic)
representation of (generalized) $\tau$-function}}}
\put(113.33,101.33){\framebox(20.33,10.00)[cc]{\parbox{.14\linewidth}{Lax
representation}}}
\put(114.33,117.67){\framebox(20.33,9.00)[cc]{\parbox{.13\linewidth}{spectral
curve}}}
\put(93.33,136.00){\framebox(19.67,9.67)[cc]
{\parbox{.13\linewidth}{polylo\-garithms}}}
\linethickness{2pt}
\put(127.67,136.00){\framebox(26.33,15.00)[cc]
{\parbox{.17\linewidth}{quasiclassical (Whitham) hierarchies}}}
\put(60.67,141.00){\framebox(22.67,10.67)[cc]{\parbox{.14\linewidth}{topological
theories}}}
\linethickness{.4pt}
\put(57.33,128.67){\framebox(33.67,9.33)[cc]{\parbox{.23\linewidth}
{(generalized) WDVV equations}}}
\put(26.67,98.67){\framebox(24.33,10.67)[cc]{\parbox{.17\linewidth}{perturbative
limit}}}
\put(32.33,117.00){\framebox(24.00,9.33)[cc]
{\parbox{.17\linewidth}{brane construction}}}
\linethickness{2pt}
\put(5.67,116.00){\framebox(16.33,9.00)[cc]{\parbox{.11\linewidth}
{SYM theories}}}
\put(0.67,73.67){\framebox(26.33,11.00)[cc]{\parbox{.18\linewidth}{string
compactifications}}}
\put(4.33,46.33){\framebox(18.33,10.00)[cc]{\parbox{.11\linewidth}{mirror
maps}}}
\linethickness{.4pt}
\put(27.33,54.00){\framebox(23.33,15.00)[cc]
{\parbox{.14\linewidth}{K3 geometry, toric geometry}}}
\put(93.67,2.33){\framebox(28.67,12.67)[cc]
{\parbox{.20\linewidth}{quantization of Dell systems}}}
\put(129.67,6.33){\framebox(25.00,15.33)[cc]
{\parbox{.17\linewidth}{quantum groups, elliptic algebras}}}
\put(92.00,121.){\makebox(0,0)[cc]{{\large Hitchin systems}}}
\emline{83.33}{147.33}{1}{127.67}{147.33}{2}
\emline{60.67}{147.00}{3}{14.00}{125.00}{4}
\emline{129.33}{136.00}{5}{90.00}{112.00}{6}
\emline{51.00}{109.33}{7}{108.67}{136.00}{8}
\emline{22.00}{116.00}{9}{26.67}{109.33}{10}
\emline{51.00}{98.33}{11}{55.67}{83.67}{12}
\emline{14.33}{116.00}{13}{14.33}{84.67}{14}
\emline{14.33}{73.67}{15}{14.33}{56.33}{16}
\emline{21.00}{73.67}{17}{27.33}{61.00}{18}
\emline{22.67}{50.00}{19}{40.00}{54.00}{20}
\emline{32.33}{117.00}{21}{6.67}{84.67}{22}
\emline{125.67}{99.83}{23}{125.67}{95.33}{24}
\emline{113.33}{81.67}{25}{92.67}{84.67}{26}
\emline{113.33}{76.00}{27}{83.33}{28.33}{28}
\emline{113.33}{60.00}{29}{92.33}{68.00}{30}
\emline{113.33}{52.67}{31}{82.33}{12.33}{32}
\emline{54.33}{16.00}{33}{64.67}{7.00}{34}
\emline{44.67}{32.33}{35}{55.67}{49.33}{36}
\emline{113.33}{34.00}{37}{82.33}{7.00}{38}
\emline{113.33}{38.00}{39}{92.67}{55.00}{40}
\emline{103.33}{15.00}{41}{84.33}{40.67}{42}
\emline{110.33}{15.00}{43}{121.67}{26.00}{44}
\emline{140.00}{26.00}{45}{142.33}{21.67}{46}
\emline{122.33}{8.00}{47}{129.67}{13.00}{48}
\emline{50.67}{62.67}{49}{64.33}{47.33}{50}
\emline{13.00}{46.33}{51}{12.95}{43.11}{52}
\emline{12.95}{43.11}{53}{13.04}{40.02}{54}
\emline{13.04}{40.02}{55}{13.27}{37.06}{56}
\emline{13.27}{37.06}{57}{13.63}{34.23}{58}
\emline{13.63}{34.23}{59}{14.14}{31.53}{60}
\emline{14.14}{31.53}{61}{14.78}{28.96}{62}
\emline{14.78}{28.96}{63}{15.56}{26.52}{64}
\emline{15.56}{26.52}{65}{16.48}{24.21}{66}
\emline{16.48}{24.21}{67}{17.53}{22.03}{68}
\emline{17.53}{22.03}{69}{18.73}{19.98}{70}
\emline{18.73}{19.98}{71}{20.06}{18.06}{72}
\emline{20.06}{18.06}{73}{21.53}{16.26}{74}
\emline{21.53}{16.26}{75}{23.14}{14.60}{76}
\emline{23.14}{14.60}{77}{24.89}{13.07}{78}
\emline{24.89}{13.07}{79}{26.77}{11.67}{80}
\emline{26.77}{11.67}{81}{28.80}{10.40}{82}
\emline{28.80}{10.40}{83}{30.96}{9.25}{84}
\emline{30.96}{9.25}{85}{33.26}{8.24}{86}
\emline{33.26}{8.24}{87}{35.70}{7.36}{88}
\emline{35.70}{7.36}{89}{38.27}{6.61}{90}
\emline{38.27}{6.61}{91}{40.99}{5.98}{92}
\emline{40.99}{5.98}{93}{43.84}{5.49}{94}
\emline{43.84}{5.49}{95}{46.83}{5.12}{96}
\emline{46.83}{5.12}{97}{49.96}{4.89}{98}
\emline{49.96}{4.89}{99}{53.23}{4.79}{100}
\emline{53.23}{4.79}{101}{56.64}{4.81}{102}
\emline{56.64}{4.81}{103}{60.18}{4.97}{104}
\emline{60.18}{4.97}{105}{64.67}{5.33}{106}
\emline{64.67}{28.33}{107}{61.34}{30.84}{108}
\emline{61.34}{30.84}{109}{58.18}{33.39}{110}
\emline{58.18}{33.39}{111}{55.21}{35.97}{112}
\emline{55.21}{35.97}{113}{52.42}{38.59}{114}
\emline{52.42}{38.59}{115}{49.81}{41.25}{116}
\emline{49.81}{41.25}{117}{47.39}{43.94}{118}
\emline{47.39}{43.94}{119}{45.14}{46.66}{120}
\emline{45.14}{46.66}{121}{43.07}{49.42}{122}
\emline{43.07}{49.42}{123}{41.19}{52.22}{124}
\emline{41.19}{52.22}{125}{39.49}{55.05}{126}
\emline{39.49}{55.05}{127}{37.96}{57.91}{128}
\emline{37.96}{57.91}{129}{36.62}{60.82}{130}
\emline{36.62}{60.82}{131}{35.46}{63.75}{132}
\emline{35.46}{63.75}{133}{34.48}{66.72}{134}
\emline{34.48}{66.72}{135}{33.68}{69.73}{136}
\emline{33.68}{69.73}{137}{33.07}{72.77}{138}
\emline{33.07}{72.77}{139}{32.63}{75.85}{140}
\emline{32.63}{75.85}{141}{32.38}{78.96}{142}
\emline{32.38}{78.96}{143}{32.30}{82.11}{144}
\emline{32.30}{82.11}{145}{32.41}{85.30}{146}
\emline{32.41}{85.30}{147}{32.70}{88.52}{148}
\emline{32.70}{88.52}{149}{33.17}{91.77}{150}
\emline{33.17}{91.77}{151}{33.82}{95.06}{152}
\emline{33.82}{95.06}{153}{34.65}{98.39}{154}
\emline{34.65}{98.39}{155}{35.66}{101.75}{156}
\emline{35.66}{101.75}{157}{36.86}{105.14}{158}
\emline{36.86}{105.14}{159}{38.23}{108.57}{160}
\emline{38.23}{108.57}{161}{39.79}{112.04}{162}
\emline{39.79}{112.04}{163}{41.53}{115.54}{164}
\emline{41.53}{115.54}{165}{43.45}{119.08}{166}
\emline{43.45}{119.08}{167}{45.55}{122.65}{168}
\emline{45.55}{122.65}{169}{47.83}{126.26}{170}
\emline{47.83}{126.26}{171}{50.29}{129.90}{172}
\emline{50.29}{129.90}{173}{52.93}{133.58}{174}
\emline{52.93}{133.58}{175}{55.76}{137.29}{176}
\emline{55.76}{137.29}{177}{60.67}{143.33}{178}
\linethickness{2pt}
\put(123.17,114.83){\thicklines\oval(31.67,29.67)[b]}
\put(124.50,114.67){\thicklines\oval(29.00,26.67)[rt]}
\thicklines\emline{107.33}{114.67}{179}{34.00}{114.67}{180}
\thicklines\emline{124.67}{128.00}{181}{34.00}{128.00}{182}
\put(34.00,121.34){\thicklines\oval(5.67,13.33)[l]}
\linethickness{.4pt}
\thinlines
\emline{22.00}{120.67}{183}{30.67}{121.00}{184}
\emline{92.67}{95.67}{185}{108.67}{102.33}{186}
\emline{127.67}{137.00}{187}{91.00}{130.33}{188}
\emline{72.00}{141.00}{189}{72.00}{138.00}{190}
\emline{22.00}{125.00}{191}{57.33}{133.67}{192}
\emline{72.33}{128.67}{193}{73.00}{111.67}{194}
\end{picture}
\caption{Around double elliptic systems}\label{gt}
\end{figure}
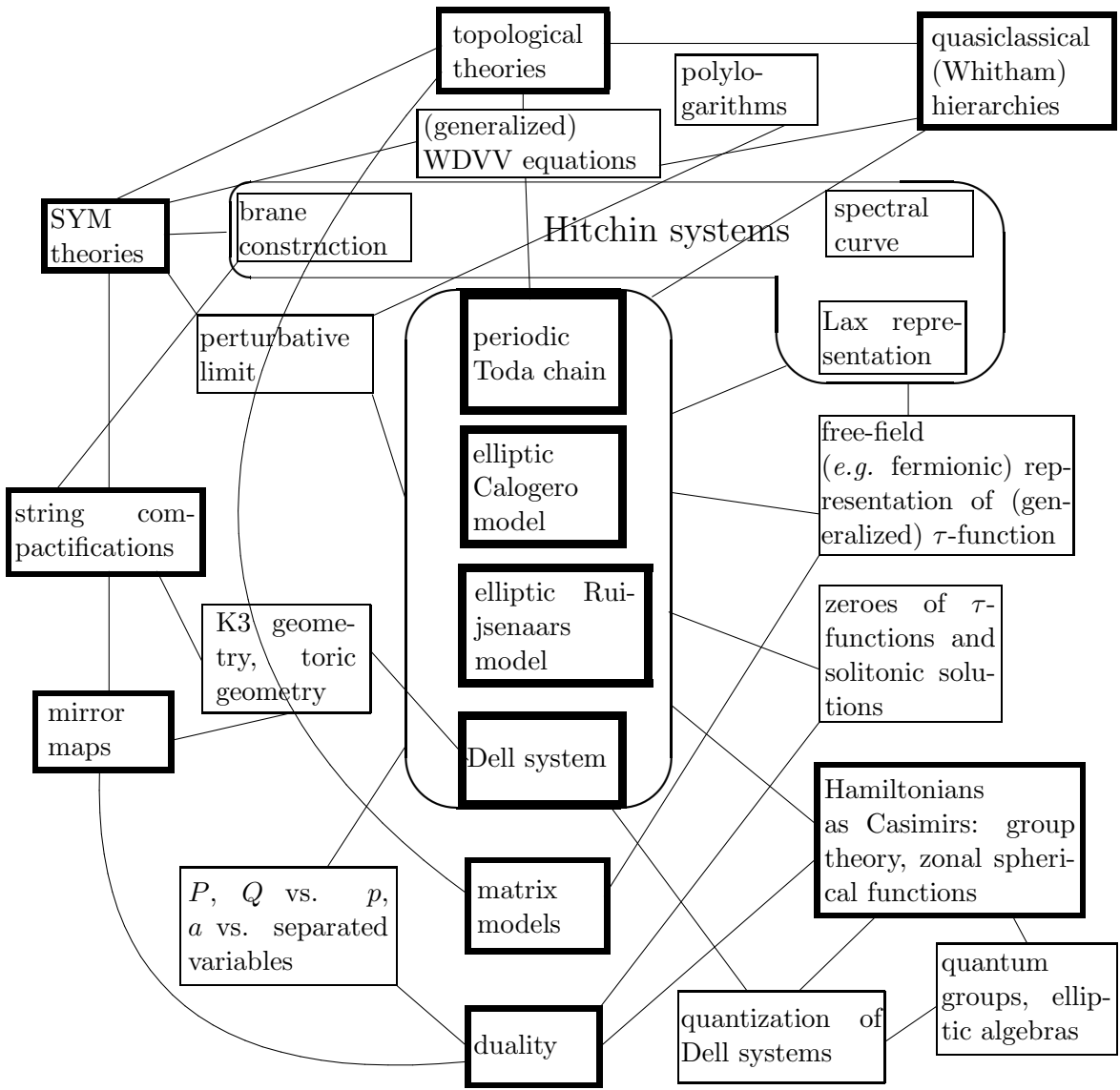

Roughly, one can divide the interest to the Dell systems into two flows:
coming from string physics and from the theory of integrable systems {\it per
se}. In Fig.\ref{gt} these subjects are listed at the l.h.s. and the r.h.s.
respectively.

\subsubsection*{2.1 Physics}

In physics, renewed interest to integrable systems is due to the discovery of
their role in quantum field theory: exact (non-perturbative) effective
actions usually exhibit integrable properties, if considered as functions of
boundary conditions (coordinates in the moduli space of vacua) and coupling
constants \cite{UFN2,UFN3,Mir93}.
The most primitive classical multiparticle integrable systems
(well reduced KP/Toda-lattice models) emerge in this way \cite{GKMMM}
in the study of vacuum dependencies of the low-energy effective actions of
$N=2$ SUSY Yang-Mills theories \cite{SW}.  The table of known relations is
shown in Fig.\ref{intvsYM}. When SUSY is broken down to $N=1$ or $N=0$,
continuous moduli normally disappear and only particular spectral
curves survive from entire Seiberg-Witten (SW) families. However, one can
still extract the dynamical information from SW theory,
restricting the answers for correlation functions to given points
in the moduli space.

\begin{figure}[t]
\begin{center} \begin{tabular}{|c|c|c|c|c|}
\hline
{\bf SUSY}&{\bf Pure gauge}&{\bf SYM theory}&{\bf SYM theory}\\
{\bf physical}&{\bf SYM theory,}&{\bf with fund.}&{\bf with adj.}\\
{\bf theory}&{\bf gauge group $G$}&{\bf matter}&{\bf matter}\\
\hline
        & inhomogeneous  & elliptic  & inhomogeneous        \\
         & periodic        & Calogero & periodic \\
{\bf 4d} & Toda chain & model & $XXX$ \\
& for the dual affine ${\hat G}^{\vee}$  &({\it trigonometric}& spin chain\\
        & ({\it non-periodic}  & {\it Calogero}& ({\it non-periodic}\\
        & {\it Toda chain})& {\it model})& {\it chain})\\
\hline
        & periodic  & elliptic & periodic   \\
        & relativistic  & Ruijsenaars&$XXZ$\\
{\bf 5d} & Toda chain  & model & spin chain  \\
        & ({\it non-periodic} & ({\it trigonometric}& ({\it non-periodic}\\
        & {\it chain}) & {\it Ruijsenaars})&{\it chain})\\
\hline
    &periodic & Dell & periodic\\
    & "Elliptic" & system & $XYZ$ (elliptic)\\
{\bf 6d} &  Toda chain & ({\it dual to elliptic} & spin chain\\
        & ({\it non-periodic}& {\it Ruijsenaars,}&({\it non-periodic} \\
& {\it chain})& {\it elliptic-trig.})&{\it chain})\\
\hline
\end{tabular}
\end{center}
\caption{
SUSY gauge theories $\Longleftrightarrow$ integrable systems
correspondence. The perturbative limit is marked by the italic font (in
parenthesis).
}\label{intvsYM}
\end{figure}
\vspace{10pt}

The most interesting distinctions between various models
in Fig.\ref{intvsYM} are:

(i) UV-finite {\it vs.} UV-infinite (where dimensional transmutation occurs)
models. In the language of integrable systems the difference is in the {\it
bare} spectral surface (complex curve):  it may be elliptic one (torus)
for the UV-finite models\footnote{The situation is still
unclear in application to the case of fundamental matter with $N_f =
2N_c$. In existing formulation for spin chains the bare coupling constant
appears rather as a twist in gluing the ends of the chain together
\cite{GGM1} (this parameter occurs only when $N_f = 2N_c$) and is not
immediately identified as a modulus of a {\it bare} elliptic curve. This
problem is a fragment of a more general puzzle:  spin chains have not been
described as Hitchin systems; only the ``$2\times 2$'' Lax representation is
known, while its ``dual'' $N_c\times N_c$ one is not yet available.}, while
it degenerates into a punctured sphere in UV-infinite cases.

(ii) Matter in adjoint {\it vs.} fundamental representations of the gauge
group.  Matter in adjoint representation can be described in terms of a
larger pure SYM model, either with higher SUSY or in higher space-time
dimension.  Thus models with adjoint matter form a hierarchy, naturally
associated with the hierarchy of integrable models {\it Toda chain
$\hookrightarrow$ Calogero $\hookrightarrow$
Ruijsenaars $\hookrightarrow$ Dell}
\cite{GKMMM,intadj,IM1,IM2,BMMM2,BMMM3}. Similarly, the
models with fundamental matter are related to the hierarchy of spin chains
originated from the Toda chain: {\it Toda chain
$\hookrightarrow$ XXX $\hookrightarrow$
XXZ $\hookrightarrow$ XYZ}
\cite{intfund,GGM1,6d}.

(iii) SYM models in different dimensions.
Actually, Yang-Mills fields play a distinguished role (representing the
universality class of conformal invariant models, perhaps, broken by the
dimensional transmutation) only in 4 and 5 dimensions.  Generically, in $d =
2k$ their role is taken by theories of $(k-1)$-forms.  The relevant model for
odd $d = 2k+1$ can be described both in terms of $(k-1)$- and $k$-forms (like
$3d$ Chern-Simons model is expressed both in terms of gauge vectors and in
terms of the scalar fields of $2d$ WZNW model).  Integrable systems relevant
for the description of vacua of $d=4$ and $d=5$ models are
respectively the Calogero and Ruijsenaars ones (which possess the ordinary
Toda chain and ``relativistic Toda chain'' as Inosemtsev's limits
\cite{Ino}).

SYM models can be embedded into the more general context of string models --
and, thus, into the entire framework of the future string theory -- in two
ways: through stringy compactifications \cite{KV} and through non-trivial
brane configurations \cite{branes}. The both approaches can be dealt with in
different ways, starting from different superstringy models in $D=10,11,12$
dimensions.  Actually, the relevant brane configurations involve branes with
some compact dimensions and, therefore, this is actually also the story about
compactifications, only from lower $D = 6,7,8$. The compact dimensions are
interpreted as complex spectral curves within the brane approach \cite{Wbr}
or their non-compact $3_Cd$ mirrors \cite{Guk} -- the singular
Calabi-Yau-like manifolds -- in string compactifications. The Lax operators
are provided by non-trivial solutions of the scalar-fields equations of
motion on the bare spectral curve \cite{MarMarM} (in other words, an attempt
to compactify on a {\it bare} spectral curve ends up with the dynamical
formation of a more sophisticated {\it full} spectral curve, {\it i.e.} the
bare curve gets "spontaneously" fibrated).

As usual, when compactification size is large as compared to the Plankian
(stringy) scale, one can consider the effective low-energy theory, neglecting
the tower of stringy massive states and gravitational interactions.  This
effective theory includes massless gauge fields and their relatively light
companions with masses inversely proportional to the compactification size,
which can be identified with the gauge fields acquiring their masses through
the Higgs mechanism and with the solitons/monopoles associated with strings
winding around the compact dimensions. The description of such effective
theory is in terms of {\it prepotentials} that celebrates a lot of
properties familiar from the original studies of pure topological theories
(where have been neglected the possibility of the light excitations to move).
These properties include the identification \cite{GKMMM,IM2,RG} of
prepotentials as quasiclassical (Whitham)
$\tau$-functions \cite{Kri,Dub} and peculiar equations, of which the
(generalized) WDVV equations \cite{WDVV,Dub,MMM} are the best known example.

In the prepotential, the contributions of particles and
solitons/mo\-no\-po\-les (dyons) sharing the same mass scale, are still
distinguishable, because of different dependencies on the bare coupling
constant, {\it i.e.} on the modulus $\tau$ of the bare coordinate elliptic
curve (in the UV-finite case) or on the $\Lambda_{QCD}$ parameter (emerging
after dimensional transmutation in UV-infinite cases).  In the limit $\tau
\rightarrow i\infty$ ($\Lambda_{QCD} \rightarrow 0$), the solitons/monopoles
do not contribute and the prepotential reduces to the ``perturbative'' one,
describing the spectrum of non-interacting {\it particles}.  It is
immediately given by the SUSY Coleman-Weinberg formula \cite{MMM}:

\be
{\cal F}_{pert}(a) =
\sum_{\hbox{\footnotesize reps}
\ R,i} (-)^F {\rm Tr}_R (a + M_i)^2\log (a + M_i)
\ee
SW theory (actually, the identification of appropriate integrable
system) can be used to construct the non-perturbative prepotential,
describing the mass spectrum of all the ``light'' (non-stringy) excitations
(including solitons/monopoles).
Switching on Whitham times \cite{RG} presumably allows one to extract
some correlation functions in the ``light'' sector.

In accordance with the general logic of string theory \cite{UFN2}, every
model of quantum field theory possesses different descriptions in terms of
different perturbative Lagrangians (referring to different
phases). Thus, the models looking different perturbatively can be, in fact,
identical non-perturbatively. If many enough couplings are
allowed, every particular model gets embedded into a unique and universal
stringy ``theory of everything''. At the present stage of knowledge we just
begin to observe particular traces of this general phenomenon in the
form of sporadic ``duality'' identities between particular string models
\cite{dualityrev}.
A further restricted class of dualities in application to $N=2$ SYM can be
identified with simple canonical transformations. This is the
coordinate-momentum duality of integrable systems \cite{duality}.

\subsubsection*{2.2 Integrability theory}

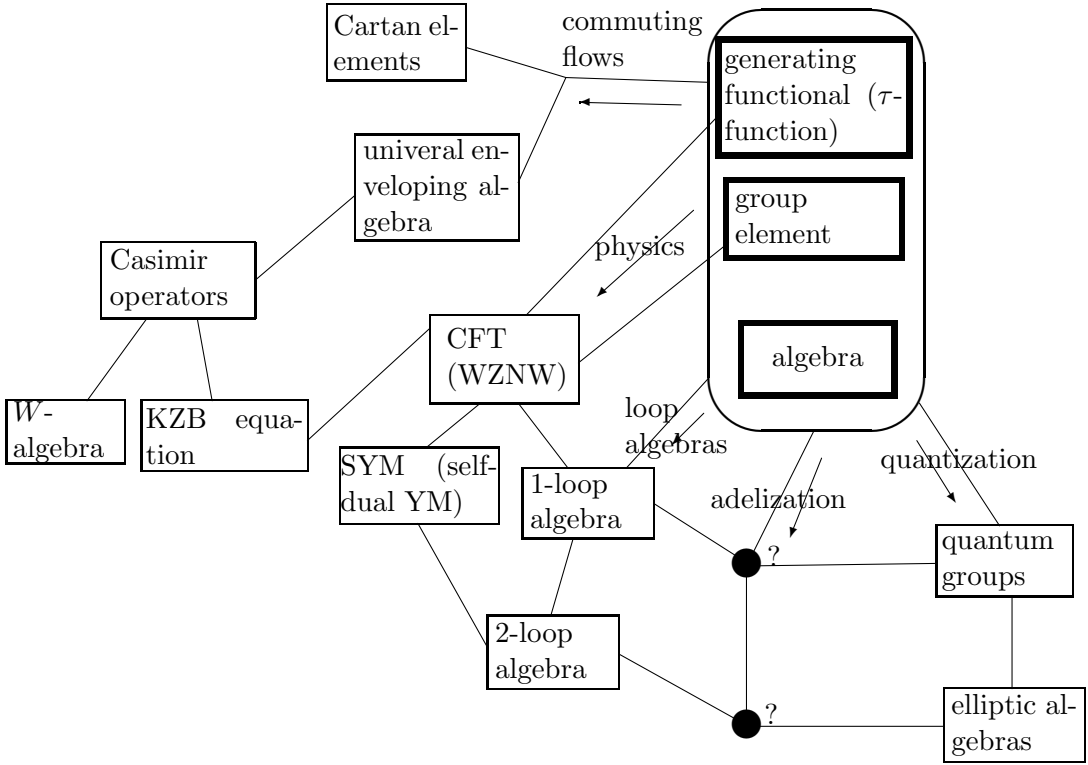
\begin{figure}[t]
\special{em:linewidth 0.4pt}
\unitlength 1.00mm
\linethickness{0.4pt}
\begin{picture}(144.00,153.00)
\linethickness{2pt}
\put(98.66,101.33){\framebox(19.67,9.00)[cc]{algebra}}
\put(96.66,119.67){\framebox(22.67,9.67)[cc]{\parbox{.15\linewidth}{group
element}}}
\put(95.66,133.33){\framebox(24.67,14.67)[cc]
{\parbox{.17\linewidth}{generating functional ($\tau$-function)}}}
\linethickness{2pt}
\put(108.33,124.33){\thicklines\oval(28.67,56.00)[]}
\linethickness{.4pt}
\put(69.33,82.00){\framebox(17.33,9.33)[cc]{\parbox{.11\linewidth}{1-loop
algebra}}}
\put(64.67,62.00){\framebox(17.33,9.67)[cc]{\parbox{.11\linewidth}{2-loop
algebra}}}
\put(124.33,74.33){\framebox(18.00,9.33)[cc]{\parbox{.12\linewidth}{quantum
groups}}}
\put(125.33,52.33){\framebox(18.67,9.67)[cc]{\parbox{.12\linewidth}{elliptic
algebras}}}
\put(99.00,78.67){\circle*{3.89}}
\put(43.33,143.00){\framebox(18.33,10.00)[cc]{\parbox{.12\linewidth}{Cartan
elements}}}
\put(47.00,121.67){\framebox(21.67,14.00)[cc]
{\parbox{.14\linewidth}{univeral enveloping algebra}}}
\put(13.33,111.33){\framebox(20.33,10.00)[cc]{\parbox{.13\linewidth}{Casimir
operators}}}
\put(0.67,92.00){\framebox(15.67,8.33)[cc]{\parbox{.10\linewidth}{$W$-algebra}}}
\put(18.66,91.00){\framebox(22.00,9.33)[cc]{\parbox{.15\linewidth}{KZB
equation}}}
\put(99.00,57.33){\circle*{4.00}}
\put(57.00,100.00){\framebox(19.67,11.67)[cc]{\parbox{.11\linewidth}
{CFT (WZNW)}}}
\put(45.00,84.00){\framebox(21.00,10.00)[cc]{\parbox{.15\linewidth}{SYM
(self-dual YM)}}}
\emline{94.00}{142.67}{1}{75.00}{143.33}{2}
\emline{75.00}{143.33}{3}{68.67}{129.33}{4}
\emline{75.00}{143.33}{5}{61.67}{147.33}{6}
\emline{47.00}{127.67}{7}{33.67}{116.33}{8}
\emline{26.00}{111.33}{9}{28.00}{100.33}{10}
\emline{19.33}{111.33}{11}{11.33}{100.33}{12}
\emline{94.00}{103.33}{13}{83.00}{91.33}{14}
\emline{68.67}{100.00}{15}{75.33}{91.33}{16}
\emline{55.33}{84.00}{17}{64.67}{67.33}{18}
\emline{76.00}{82.00}{19}{73.00}{71.67}{20}
\emline{86.67}{86.67}{21}{98.67}{79.00}{22}
\emline{82.00}{66.67}{23}{99.33}{57.00}{24}
\emline{108.00}{96.33}{25}{99.00}{78.33}{26}
\emline{99.00}{78.33}{27}{99.00}{57.00}{28}
\emline{99.00}{78.33}{29}{124.33}{78.67}{30}
\emline{99.00}{57.00}{31}{125.33}{57.00}{32}
\emline{122.00}{100.00}{33}{132.67}{83.67}{34}
\emline{134.33}{74.33}{35}{134.33}{62.00}{36}
\put(93.33,96.67){\makebox(0,0)[cc]{\parbox{.15\linewidth}{loop algebras}}}
\put(103.33,87.33){\makebox(0,0)[cc]{adelization}}
\put(127.33,92.33){\makebox(0,0)[cc]{quantization}}
\put(84.33,148.50){\makebox(0,0)[cc]{\parbox{.14\linewidth}{commuting flows}}}
\put(89.33,120.67){\makebox(0,0)[cc]{\parbox{.15\linewidth}{physics}}}
\put(89.33,94.67){\vector(-1,-1){0.2}}
\emline{93.33}{98.67}{37}{89.33}{94.67}{38}
\put(77.00,140.00){\vector(-1,0){0.2}}
\emline{90.33}{139.67}{39}{77.00}{140.00}{40}
\put(105.00,82.67){\vector(-1,-3){0.2}}
\emline{109.00}{92.67}{41}{105.00}{82.67}{42}
\put(126.67,87.00){\vector(2,-3){0.2}}
\emline{121.67}{95.00}{43}{126.67}{87.00}{44}
\emline{40.66}{95.00}{45}{57.00}{110.00}{46}
\emline{95.67}{138.67}{47}{69.67}{111.67}{48}
\emline{96.67}{121.33}{49}{76.67}{105.33}{50}
\emline{63.67}{100.00}{51}{56.33}{94.00}{52}
\put(79.33,114.33){\vector(-1,-1){0.2}}
\emline{92.00}{125.67}{53}{79.33}{114.33}{54}
\put(102.67,79.67){\makebox(0,0)[cc]{?}}
\put(102.33,58.67){\makebox(0,0)[cc]{?}}
\end{picture}
\vspace{-5.5cm}
\caption{Some relevant branches of group theory}\label{group}
\end{figure}

Integrability theory studies commuting Hamiltonian flows.
A minor generalization allows the flows to form a closed non-abelian algebra.
Thus, integrability theory is naturally a branch of group theory (see
Fig.\ref{group}),
studying the group element of a (Lie, quantum, elliptic, ...) algebra
as a (classical, quantum, elliptic(?), ...) multi-time evolution operator.
The central object in the theory is the collection of matrix elements
of the evolution operator in a given representation assembled into
a generating function called (generalized) $\tau$-function \cite{UFN3,Mir98}
of the corresponding algebra. Morphisms of representations
(like decomposition of tensor products $R_i\otimes R_j =
\oplus_k N_{ij}^k R_k$) induce Hirota-like relations between the
$\tau$-functions in different representations \cite{GKLMM},
which can be rewritten as a hierarchy of
(differential, finite-difference, elliptic(?)...) equations.

\vspace{10pt}

From this perspective, integrability is not so much concentrated on
Hamiltonians and their commutativity, but the really important issue becomes
links with representation theory. Still, if one is interested in {\it
commuting} flows, the two natural places for
the commuting flows to emerge are in studies of the Cartan-like subalgebras
and of the Casimir operators. If concentrated on Hamiltonians, people often
distinguish between the classical Hamiltonian equations of motion, ${\partial
p_i\over\partial t_n}=-{\partial H_n\over\partial q_i}$, ${\partial
q_i\over\partial t_n}={\partial H_n\over\partial p_i}$ and the
Schr\"odinger equations $i{\partial\Psi\over\partial t_n}=\hat H_n\Psi$.
Then, a way from group theory to integrable Hamiltonians
is through restricting Casimir operators to particular orbits
(the theory of Hamiltonian reductions \cite{Hred}). This gives rise to
Laplace-like (differential and shift) operators together with their
eigenfunctions (wave functions or zonal spherical functions). The
eigenfunctions being represented as particular matrix elements are contained
in the (generalized) $\tau$-function. The problem here is to extend the
formalism to more sophisticated groups, like quantum affine \cite{ChP} and
elliptic \cite{ellalg} algebras. Only at affine elliptic level one expects the
Dell hamiltonians to appear.  Note that the coordinate-momentum duality acts
on the zonal spherical functions, interchanging the arguments and
eigenvalues \cite{qduality}.

An {\it a priori} different {\it quantization} of integrability comes
from a different place:  classical and quantum integrability can be
associated with continuous and discrete-time evolutions and thus related to
classical and quantum groups respectively.  The most intriguing from the
group theory point of view is the quasiclassical (Whitham) integrability
\cite{Kri,EdinMor}. Further generalization to the case of a two-component
Plank constant imply a sort of elliptic quantization associated with emerging
elliptic algebras \cite{ellalg}. Their full description will hopefully shed
light also on the somewhat mysterious ``p-adic dimension'', empirically
observed, among other places, in the properties of McDonald polynomials
\cite{FZ}.

On the other hand, elliptic algebras are naturally expected to appear
\cite{GLM} in the study of loop algebras ({\it cf.} with Kazhdan-L\"ustig
framework, see also \cite{Cher}):  given an algebra, one can consider the
corresponding 1-loop, 2-loop etc algebras (mapping from circles, Riemann
surfaces etc into the algebra).  At every level, an interesting new
deformation shows up. At the 1-loop level, the central extensions give rise
to affine (Kac-Moody) algebras. At the 2-loop level, a Moyal bracket (quantum
deformation) should occur etc.

The most adequate tool for group theory investigations is provided by quantum
field theory (geometrical quantization). The hidden group-theoretical
structures manifest themselves by high quality of the approximation, provided
by the transformation to free fields (Darboux variables). ``High quality''
normally means that the corrections are concentrated in co-dimension one or
higher (in other words, the original theory can be substituted by a mixture
of the free-field ones with various non-trivial boundary conditions, {\it
i.e.} with non-trivial theories on the boundaries). To put it differently,
the functional integral for $d$-dimensional theory in compact space-time can
be reduced to a collection of $d$-dimensional Green functions (correlators of
free fields) and non-trivial functional integral over fields on some
hypersurfaces. This phenomenon is well studied (and partly understood) in the
theory of $2d$ conformal models: the WZNW model can be represented in terms
of free fields with insertions of additional "screenings" realized as contour
integrals \cite{GMMOS}. Another (not unrelated) example is provided by the
old theory of hierarchy of anomalies \cite{UFNan}.  The hierarchy relates,
for example, the Donaldson theory in $4d$ (3-loop level), Chern-Simons theory
in $3d$ (2-loop level; co-dimension one), WZNW theory in $2d$ (1-loop level;
co-dimension two) etc.

The free field formulation is intensively used in the
theory of conventional KP/Toda-lattice $\tau$-functions \cite{DJKM}
(of which the Dell-Ruijsenaars-Calogero-Toda-chain family is a
very specific reduction)
associated with the affine (Kac-Moody) algebras of level $k=1$
and also with the {\it eigenvalue} matrix models \cite{UFN3,Mir93}.
The next step will be made when more general ``non-abelian''
or ``non-eigenvalue'' $\tau$-functions associated with
the level $k\neq 1$ emerge from the future studies of effective
actions (generating functions of correlators) of the $2d$ WZNW
model and $3d$ Chern-Simons theory.
However, the real breakthrough, when the interrelation
between different fashionable subjects is to become transparent,
is expected after the $\tau$-functions of double-loop algebras
(like that of area-preserving diffeomorphisms and/or the ones
implicit in emerging the deformation quantization theory \cite{Kon})
get enough attention.

In algebraic geometry, instead of studying representation theory of
groups {\it per se}, in Tanaka-Krein style,
one considers particular sophisticated representations, like those
in bundles/sheaves over complex curves and surfaces, thus
mixing the group theory and algebraic geometry data.
This line of research is nowadays labeled as Hitchin \cite{Hitchin} or
SW theory and it provides the most of currently
intriguing links to non-perturbative dynamics of quantum physical
systems. Remarkably, it is enough to input just very simple
{\it bare} spectral curves, like punctured Riemann sphere or torus, in order
to get from the geometrical quantization technique some very non-trivial
families of the {\it full} spectral curves
(dynamics induced by group theory "spontaneously" transforms
trivial curves into highly non-trivial ones)
and somewhat sophisticated prepotentials belonging to the fashionable
classes of special functions (like polylogarithims and
their generalizations).
The most interesting issue here is that the prepotential
turns out to be a quasiclassical $\tau$-function \cite{GKMMM,RG}
and, therefore, it can be obtained in two very different ways:
from SW/Hitchin theory, where, at least, the {\it bare}
spectral curves are used as input, and from quasiclassical
limits of ordinary (generalized) $\tau$-functions built directly
from affine quantum groups, where the only naive origin
for a {\it curve} to appear is due to dealing with loop algebras.
This situation echoes the general one with the quasiclassical quantization of
stringy theory. Quasiclassical approximation always depends on two things:
dynamics itself (which is dictated by symmetry principles and group
theory) and the choice of ``vacuum'' (which usually breaks a lot
of symmetries and is naturally an object of algebraic geometry nature).
Thus, quasiclassical dynamics can be described in two languages:
in terms of the full quantum dynamics ({\it i.e.} -- at the end of the day --
in terms of ordinary generalized $\tau$-functions) and in terms of
deformations of the vacuum ({\it i.e.} in terms of deformation theory,
say, of a Hodge style one). In the last description, one expects also
additional (duality) relations between quasiclassical theories nearby
different vacua to occur, which reflects the existence of the full quantum
dynamics {\it common} for all the vacua. One should keep in
mind that, in the situation with many deformation parameters,
the same words can be said about ``the quasiclassical approximation''
w.r.t. any of the parameters. The future string theory is expected to
unify in this manner all quantum field models, dealing with them as
perturbative ( = quasiclassical) approximations to its different
phases ( = vacua) with the Plank constant substituted by the inverse Plank
mass as the deformation parameter...
Coming back to SW theory, what really remains to
discover within this context is the rich structure of the ``vacua''
(associated with the variety of complex spectral curves) in the
geometrical quantization of affine quantum groups.

\subsection*{3\hspace{2mm} Open problems involving Dell systems}

Actually, most of the presentation of the previous section describe
the open problems: every line calls for further clarification
and details. Still, it makes sense to list a number of lower
scale particular problems, directly involving the Dell systems.

\paragraph{Complete construction of Dell integrable systems}
\begin{itemize}
\item
Explicit construction of Dell Hamiltonians for groups larger than
$SU(2)$
\item
Commutativity of the Hamiltonians
\item
Quantization of the Hamiltonians
\item
Inozemtsev's limit (for coordinate torus) \cite{Ino,IM1}
of the Dell system: ``elliptic Toda chain'' \cite{ellToda}
\end{itemize}
\paragraph{Identification of the relevant gauge theory}
\begin{itemize}
\item
Perturbative prepotential
\item
Relation to spin chains
\item
String compactifications
\item
Torus fibered over torus, particular K3 surfaces
\item
Relevant brane configurations and gauge theory on the brane
\item
Duality between $6d$ self-dual 2-forms and SYM model
\item
The physical meaning of the spectral manifold (curve)
and bare spectral curve {\it per se} -- without reference
to brane configuration
\end{itemize}
\paragraph{Hitchin systems}
\begin{itemize}
\item
Spectral curve
\item
SW differential
\item
Lax operator
\item
Brane configuration
\end{itemize}
\paragraph{Relation to conventional $\tau$-function}
\begin{itemize}
\item
The general problem: Hitchin {\it vs.} $\tau$
\item
Fermionic representation
\item
Free field representations
\end{itemize}
\paragraph{Dell from Casimir operators}
\begin{itemize}
\item
Quantization of integrable Hamiltonians
\item
Identification of the relevant groups, Casimir
functions, orbits, reductions
\item
Elliptic algebras
\item
Adelization
\end{itemize}
\paragraph{Duality}
\begin{itemize}
\item
Formulation of the coordinate-momentum duality, relation to
generic canonical transformations -- see Fig.\ref{dual}
\item
Duality transforms of the wave functions (zonal spherical functions)
\item
The study of the duals to the elliptic Calogero and Ruijsenaars
models (the elliptic-rational and elliptic-trigonometric systems)
\item
$p,q$ {\it vs.} $p^{Jac},a$ {\it vs.} "separated" variables ${\cal P}$,
${\cal Q}$
\item
Explicit construction of Dell systems for $SU(N)$ and other groups
\item
Commuting $\theta$-functions \cite{MM}
\item
Dual systems \cite{B,BMMM3,MM} from zeroes of $\tau$-functions \cite{zertau}
\item
Transformation of prepotentials under duality
\item
Relation to the mirror symmetry
\item
Relation to Langlands correspondence
\end{itemize}

\begin{figure}[t]
\special{em:linewidth 0.4pt}
\unitlength 1mm
\linethickness{0.4pt}
\begin{picture}(141.33,120.33)
\emline{18.33}{110.00}{1}{141.33}{110.00}{2}
\emline{141.33}{110.00}{3}{141.33}{110.00}{4}
\emline{18.33}{95.00}{5}{141.33}{95.00}{6}
\emline{18.33}{80.00}{7}{141.33}{80.00}{8}
\emline{18.33}{65.00}{9}{141.00}{65.00}{10}
\emline{141.00}{120.00}{11}{141.00}{65.00}{12}
\emline{110.33}{120.00}{13}{110.33}{65.00}{14}
\emline{80.00}{120.33}{15}{80.00}{65.00}{16}
\emline{50.00}{120.00}{17}{50.00}{65.00}{18}
\emline{50.00}{110.00}{19}{18.33}{120.00}{20}
\put(84.00,101.33){\vector(4,3){0.2}}
\emline{70.67}{91.67}{21}{84.00}{101.33}{22}
\put(73.33,90.33){\vector(-4,-3){0.2}}
\emline{85.67}{99.33}{23}{73.33}{90.33}{24}
\put(114.00,85.33){\vector(4,3){0.2}}
\emline{102.67}{76.33}{25}{114.00}{85.33}{26}
\put(105.67,75.67){\vector(-4,-3){0.2}}
\emline{115.33}{83.33}{27}{105.67}{75.67}{28}
\put(58.33,105.33){\vector(2,-1){0.2}}
\emline{54.67}{102.67}{29}{53.40}{104.44}{30}
\emline{53.40}{104.44}{31}{52.52}{105.86}{32}
\emline{52.52}{105.86}{33}{52.02}{106.94}{34}
\emline{52.02}{106.94}{35}{51.90}{107.67}{36}
\emline{51.90}{107.67}{37}{52.17}{108.05}{38}
\emline{52.17}{108.05}{39}{52.82}{108.08}{40}
\emline{52.82}{108.08}{41}{53.85}{107.77}{42}
\emline{53.85}{107.77}{43}{55.27}{107.11}{44}
\emline{55.27}{107.11}{45}{58.33}{105.33}{46}
\put(88.67,91.33){\vector(2,-1){0.2}}
\emline{84.33}{87.67}{47}{83.03}{89.38}{48}
\emline{83.03}{89.38}{49}{82.07}{90.81}{50}
\emline{82.07}{90.81}{51}{81.46}{91.96}{52}
\emline{81.46}{91.96}{53}{81.20}{92.83}{54}
\emline{81.20}{92.83}{55}{81.28}{93.43}{56}
\emline{81.28}{93.43}{57}{81.72}{93.74}{58}
\emline{81.72}{93.74}{59}{82.50}{93.77}{60}
\emline{82.50}{93.77}{61}{83.64}{93.52}{62}
\emline{83.64}{93.52}{63}{85.12}{92.99}{64}
\emline{85.12}{92.99}{65}{88.67}{91.33}{66}
\put(117.33,76.67){\vector(2,-1){0.2}}
\emline{114.00}{73.33}{67}{112.81}{75.06}{68}
\emline{112.81}{75.06}{69}{112.08}{76.42}{70}
\emline{112.08}{76.42}{71}{111.81}{77.40}{72}
\emline{111.81}{77.40}{73}{112.00}{78.00}{74}
\emline{112.00}{78.00}{75}{112.65}{78.23}{76}
\emline{112.65}{78.23}{77}{113.75}{78.08}{78}
\emline{113.75}{78.08}{79}{115.31}{77.56}{80}
\emline{115.31}{77.56}{81}{117.33}{76.67}{82}
\put(114.67,98.67){\vector(1,4){0.2}}
\emline{74.00}{76.67}{83}{77.17}{76.81}{84}
\emline{77.17}{76.81}{85}{80.21}{77.04}{86}
\emline{80.21}{77.04}{87}{83.13}{77.36}{88}
\emline{83.13}{77.36}{89}{85.92}{77.77}{90}
\emline{85.92}{77.77}{91}{88.60}{78.27}{92}
\emline{88.60}{78.27}{93}{91.15}{78.86}{94}
\emline{91.15}{78.86}{95}{93.57}{79.54}{96}
\emline{93.57}{79.54}{97}{95.88}{80.31}{98}
\emline{95.88}{80.31}{99}{98.06}{81.17}{100}
\emline{98.06}{81.17}{101}{100.11}{82.12}{102}
\emline{100.11}{82.12}{103}{102.05}{83.16}{104}
\emline{102.05}{83.16}{105}{103.86}{84.29}{106}
\emline{103.86}{84.29}{107}{105.55}{85.51}{108}
\emline{105.55}{85.51}{109}{107.11}{86.82}{110}
\emline{107.11}{86.82}{111}{108.55}{88.22}{112}
\emline{108.55}{88.22}{113}{109.87}{89.72}{114}
\emline{109.87}{89.72}{115}{111.07}{91.30}{116}
\emline{111.07}{91.30}{117}{112.14}{92.97}{118}
\emline{112.14}{92.97}{119}{113.09}{94.73}{120}
\emline{113.09}{94.73}{121}{113.92}{96.58}{122}
\emline{113.92}{96.58}{123}{114.67}{98.67}{124}
\put(76.00,73.33){\vector(-1,0){0.2}}
\emline{117.00}{96.67}{125}{116.05}{94.64}{126}
\emline{116.05}{94.64}{127}{115.01}{92.70}{128}
\emline{115.01}{92.70}{129}{113.86}{90.85}{130}
\emline{113.86}{90.85}{131}{112.60}{89.10}{132}
\emline{112.60}{89.10}{133}{111.25}{87.44}{134}
\emline{111.25}{87.44}{135}{109.79}{85.87}{136}
\emline{109.79}{85.87}{137}{108.23}{84.39}{138}
\emline{108.23}{84.39}{139}{106.57}{83.00}{140}
\emline{106.57}{83.00}{141}{104.81}{81.71}{142}
\emline{104.81}{81.71}{143}{102.94}{80.50}{144}
\emline{102.94}{80.50}{145}{100.97}{79.39}{146}
\emline{100.97}{79.39}{147}{98.90}{78.37}{148}
\emline{98.90}{78.37}{149}{96.72}{77.44}{150}
\emline{96.72}{77.44}{151}{94.45}{76.60}{152}
\emline{94.45}{76.60}{153}{92.07}{75.86}{154}
\emline{92.07}{75.86}{155}{89.58}{75.21}{156}
\emline{89.58}{75.21}{157}{87.00}{74.64}{158}
\emline{87.00}{74.64}{159}{84.31}{74.17}{160}
\emline{84.31}{74.17}{161}{81.53}{73.80}{162}
\emline{81.53}{73.80}{163}{78.63}{73.51}{164}
\emline{78.63}{73.51}{165}{76.00}{73.33}{166}
\put(39.33,119.00){\makebox(0,0)[cc]{{coordinate}}}
\put(26.67,113.33){\makebox(0,0)[cc]{{momentum}}}
\put(64.67,114.33){\makebox(0,0)[cc]{rational}}
\put(95.00,113.67){\makebox(0,0)[cc]{trigonometric}}
\put(125.00,114.00){\makebox(0,0)[cc]{elliptic}}
\put(33.67,102.00){\makebox(0,0)[cc]{rational}}
\put(33.33,86.67){\makebox(0,0)[cc]{trigonometric}}
\put(33.33,72.67){\makebox(0,0)[cc]{elliptic}}
\put(73.00,102.00){\makebox(0,0)[cc]{\parbox{.2\linewidth}{rational
Calogero}}}
\put(98.67,102.00){\makebox(0,0)[cc]{\parbox{.2\linewidth}
{trigonometric Calogero}}}
\put(130.67,102.33){\makebox(0,0)[cc]{\parbox{.15\linewidth}{elliptic
Calogero}}}
\put(65.67,86.67){\makebox(0,0)[cc]{\parbox{.2\linewidth}{rational
Ruijsenaars}}}
\put(95.67,86.33){\makebox(0,0)[cc]{\parbox{.2\linewidth}
{trigonometric Ruijsenaars}}}
\put(129.67,86.33){\makebox(0,0)[cc]{\parbox{.2\linewidth}
{elliptic Ruijsenaars}}}
\put(64.67,72.67){\makebox(0,0)[cc]{\parbox{.2\linewidth}
{dual Calogero}}}
\put(98.67,71.00){\makebox(0,0)[cc]{\parbox{.15\linewidth}
{dual Ruijsenaars}}}
\put(129.67,73.00){\makebox(0,0)[cc]{\parbox{.2\linewidth}{Dell system}}}
\end{picture}
\vspace{-6.8cm}
\caption{Action of the coordinate-momentum duality on the
Calogero-Ruijsenaars-Dell family. Hooked arrows mark self-dual
systems. The duality leaves the coupling constant $g$ intact.}\label{dual}
\end{figure}
\vspace{10pt}

\paragraph{Theory of prepotentials (quasiclassical $\tau$-functions)}
\begin{itemize}
\item
Definition in internal terms
\item
The meaning of the (generalized) WDVV equations \cite{EdinMir}
\item
Further generalization of WDVV equations
\item
RG flows as Whitham dynamics \cite{RG}
\item
Relation to (generalized) $\tau$-functions, {\it i.e.} the pure group-theory
objects
\item
Prepotentials from quantum affine algebras
\item
Relation to polylogarithms
\end{itemize}

\bigskip

\subsection*{Acknowledgments}

We are indebted for numerous discussions to H.W.Braden, A.Gerasimov, A.Gorsky,
S.Kharchev, A.Marshakov, M.Olshanetsy and A.Zabrodin.  Our work is partly
supported by the RFBR grants 98-01-00328 (A.Mir.), 98-02-16575 (A.Mor.), the
Russian President's Grant 96-15-96939, the INTAS grant 97-0103 and the program
for support of the scientific schools 96-15-96798. A.Mir. also acknowledges
the Royal Society for support under a joint project.

\end{document}